\documentclass[11pt,reqno]{article}

\usepackage{times}
\usepackage{hyperref}
\usepackage{xcolor}	
\usepackage{mathrsfs}
\definecolor{darkblue}{rgb}{0, 0, 0.5}
\hypersetup{colorlinks=true,citecolor=darkblue, linkcolor=darkblue, urlcolor=darkblue}

\usepackage{amsmath}
\usepackage{graphicx}
\usepackage{tikz}
\usepackage{amssymb}
\usetikzlibrary{arrows}
\usepackage{hyperref} 
\usepackage{amsthm}
\usepackage{float}
\usepackage{setspace}
\usepackage{authblk}
\usepackage[round]{natbib}
\usepackage{wrapfig}

\makeatletter

      \theoremstyle{plain}
      
        \newtheorem{remark}{Remark}
      \newtheorem{theorem}{Theorem}
       
\newcommand\@erelb@r[1]{%
  \mathrel{\tikz[baseline=-.5ex]\draw[#1] (0,0)--(0.3,0);}
}

\newcommand{\erelbar}[1]{\@erelbar#1}
\def\@erelbar#1#2{%
  \ifcase\numexpr#1*4+#2\relax
    \@erelb@r{-}\or     
    \@erelb@r{->}\or    
    \@erelb@r{-|}\or    
    \@erelb@r{->|}\or   
    \@erelb@r{<-}\or    
    \@erelb@r{<->}\or   
    \@erelb@r{<-|}\or   
    \@erelb@r{<->}\or   
    \@erelb@r{|-}\or    
    \@erelb@r{|->}\or   
    \@erelb@r{|-|}\or   
    \@erelb@r{|<->|}\or 
    \@erelb@r{|<-}\or   
    \@erelb@r{|<->}\or  
    \@erelb@r{|<-|}\or  
    \@erelb@r{|<->|}    
  \else
    \@wrong
  \fi
}
\makeatother

\newcommand{\gop}{
\overset{P}{\to}
}

\title{Optimality-based reward learning with applications to toxicology}

\begin{document}

\author[1]{Samuel J. Weisenthal \thanks{samuel\_weisenthal@urmc.rochester.edu}}
\author[2]{Matthew Eckard }
\author[1]{Askhan Ertefaie}
\author[1]{Marissa Sobolewski}
\author[1]{Sally W. Thurston}

\affil[1]{University of Rochester Medical Center,
    601 Elmwood Ave, 
            Rochester,
            NY,
             14642, 
            USA}

\affil[2]{Radford University,
    801 E Main St, Radford, VA, 24142, USA}
\date{}


    


\small
\maketitle

\small


\section{Abstract}
  In toxicology research,  experiments are often conducted to determine the effect of {toxicant} exposure on the behavior of mice, {where mice are randomized to receive the toxicant or not}. In particular, in fixed interval experiments, one provides a mouse reinforcers (e.g., a food pellet), contingent upon some action taken by the mouse (e.g., a press of a lever), but the reinforcers are only provided after fixed time intervals.  Often, to analyze fixed interval experiments, one specifies and then estimates the conditional state-action distribution (e.g., using an ANOVA). This existing approach, which in the reinforcement learning framework would be called modeling the mouse's ``behavioral policy,'' is sensitive to misspecification.  It is likely that any model for the behavioral policy is misspecified; a mapping from a mouse's exposure to their actions can be highly complex. In this work, we avoid specifying the behavioral policy by instead learning the mouse's reward function.  Specifying a reward function is as challenging as specifying a behavioral policy, but we propose a novel approach that incorporates knowledge of the optimal behavior, which is often known to the experimenter, to avoid specifying the reward function itself. In particular, we define the reward as a divergence of the mouse's actions from optimality, where the representations of the action and optimality can be arbitrarily complex. The parameters of the reward function then serve as a measure of the mouse's tolerance for divergence from optimality, which is a novel summary of the impact of the exposure. The parameter itself is scalar, and the proposed objective function is differentiable, allowing us to benefit from typical results on consistency of parametric estimators while making very few assumptions. The proposed method may be applicable to other settings in which behavior is observed, and some notion of optimality is available,  including variations of the fixed interval design and more complex studies, such as intermittent reinforcement experiments.

\section{Introduction}

We are often exposed to various  environmental chemicals, many of which are generated by human activity.  Many of these chemicals have uncertain effects on our health, but some are suspected to be toxic even in relatively small  doses \citep{gilbert2004small}. Legislation controlling the presence of certain {toxicant}s in the environment is often based on scientific studies.  However, it is challenging to conduct such studies using human subjects. Animal models, however, can help us better understand how chemicals affect biological organisms, in turn allowing us to better understand the effect of these chemicals on humans, and, usually in conjunction with observational studies, strengthening the case that the levels of certain chemicals should be regulated.  

\begin{figure}[htp!]
        \centering
    \includegraphics[width=0.5\textwidth]{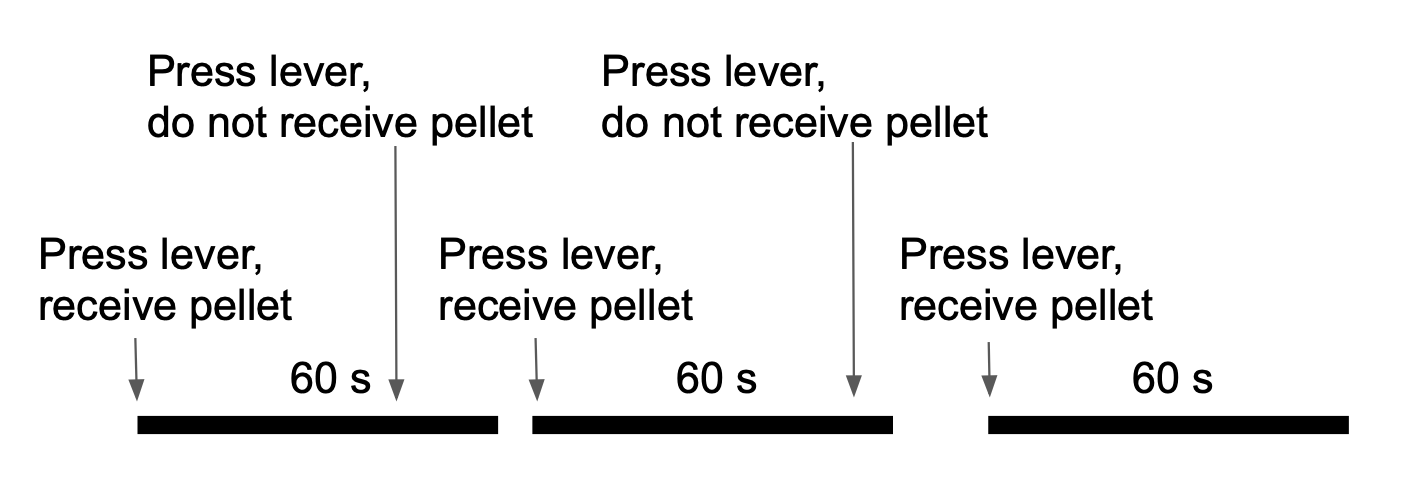}
    \caption[Fixed interval experiment]{A fixed interval experiment { showing receipt of reward pellets in relation to the timing of lever presses by the mouse.}
    }
     
    \label{fig:fixedIntervalExperiment}
    \end{figure}

In toxicology,  mice may be randomized to receive  a particular dose of a toxicant and their subsequent actions measured (see, e.g., \citet{ferster1957schedules,schneider1969two,sobolewski2014,eckard2023neonatal}). In a fixed interval experiment, which we analyze here, a mouse  can take some action (e.g., pressing a lever), and in turn receive some reinforcer, such as a pellet of food.  Receiving the food pellet begins a 60 s fixed interval, and if the mouse takes the action again before the fixed interval has elapsed, they receive no further reinforcer.   However, if they take an action after the 60 s have elapsed,  they receive another reinforcer, the reception of which begins the next fixed interval. In Figure \ref{fig:fixedIntervalExperiment}, we show an example of  three 60 s intervals in a fixed interval experiment, where the reinforcer is a food pellet and the action is the press of a lever.   The fixed interval experiments we consider in this study occur within a ``session'' of length 30 minutes, so the mouse has a 30 minute block in which it can press as often as desired, receiving up to 30 food pellets and initiating up to 30 fixed intervals.

Usually, fixed interval experiments are analyzed by estimating the parameters that govern the animal’s behavior, where the animal's behavior is reflected by the actions it takes.  For example, one models the rate at which the animal pushes the lever (the action), with covariates corresponding  to the different exposures (the states).  In fact, this model is, for a conditional distribution of actions given states, called the ``behavioral policy'' in the reinforcement learning literature \citep{sutton2018reinforcement}. If we consider our outcome to be the summaries of the animal actions, such as the mean rate of lever presses, then modeling the behavioral policy is equivalent to an Analysis of Variance (ANOVA) \citep{scheffe1999analysis}, which is therefore commonly used for these types of experiments (often, a repeated measures ANOVA is used to assess behavior over time).  An ANOVA might show that some {toxicant}s increase a summary statistic of the animal’s response rate, as evidenced by a large coefficient on the exposure variable.  Indeed, under certain exposures, such as 2,3,7,8-Tetrachlorodibenzodioxin (TCDD), mice show an inability to inhibit responding when working for the reinforcer, and might continue to press the lever many times before the 60 s has elapsed, even when receiving no food pellet \citep{sobolewski2014}. 
A major difficulty, however, is in correctly specifying the model for the animal's behavior. The mapping from exposure, {which is often received in utero,} to a mouse's behavior is likely highly complex, {and it is unlikely that the mapping from this exposure to the action in a complex behavioral experiment such as a fixed interval design corresponds to the model in an ANOVA}.  If the behavioral policy model proposed by the investigator is incorrect, the estimate that is obtained does not correspond to a true estimand reflecting the {toxicant}'s effect on the animal's behavior.  

The proposed approach,  which we will now describe,allows us to sidestep having to specify the behavioral policy. Rather, we model the animal's reward function.   One can define a reward function for a mouse as a measure of the utility that the animal achieves after taking particular actions.  For example, if the mouse times its presses well, it might receive 28 food pellets and only press once for each pellet, receiving many food pellets and also conserving energy.  This will lead to a high reward.  If a mouse however presses too frequently, it might also receive 28 food pellets, but it might expend more energy than necessary, leading to a low reward. Also, a mouse might conserve energy, only pressing a few times, but it might only receive a few food pellets, leading to a low reward overall. One can formally imagine a reward as a ``function'' of the animal's actions, the number of food pellets received in a session, and the number of calories expended.
 { Such a reward is generally as difficult or more difficult to specify than a behavioral policy, but, in this work, we do not model the reward directly.}

Our work centers around the idea that the scientist who designed the fixed interval experiment has  knowledge about what constitutes ``optimal'' behavior.  For example, in the experiment described above, the experimenter knows that it is better for the mouse to wait and press the lever after 60 s rather than to expend energy with futile presses before the 60 s have elapsed. Sometimes, investigators see recurring, sophisticated optimal behaviors, such as scalloping, in which the animal's response rates ramp up around the  end of the fixed interval \citep{skinner1938behavior}. 
Because the experimenter {might have a sense, often based on observed behavior of mice during their terminal sessions, of} what the optimal action is, we can {avoid specifying the reward function directly, and instead} define a reward function in terms of the divergence from this {proposed} optimality. The tolerance for divergence from optimality then serves as a novel summary statistic,    which can be derived for the exposed and unexposed mice (or for each dose of the exposure) and then serves as a novel summary of the effect of the exposure.  Modeling the tolerance for divergence from optimality only requires choosing a way to quantify the divergence.  More importantly, the estimand in our case is always well defined as a tolerance for divergence from optimality, whereas in the case of a misspecified behavioral policy, the estimand of the specified model may not correspond to any true estimand. {Further, in the proposed framework, one can represent the mouse's action and optimality in arbitrarily complex ways, allowing one to model complex experiments and incorporate sophisticated definitions of optimality.  Finally, the proposed estimand is a scalar, and the objective function and framework are completely parametric, allowing one to use parametric theory to prove properties such as consistency, while still making very few assumptions about the nature of the distributions involved.}  

{There is a rich body of work on modeling the timing of the animal's lever presses in fixed interval experiments \citep{gibbon1977scalar,machado1997learning}. Although these approaches provide interesting internal models of animal behavior, they do not cast it within a decision-theoretic, reinforcement learning framework.}  Our proposed approach falls in line with a body of literature on using the reinforcement learning framework to model behavior \citep{niv2009reinforcement}.   In particular, our method is a novel approach to inverse-reinforcement learning \citep{russell1998learning}, a growing research area in which one attempts to learn a reward function from data.  Although there has been  some activity on analyzing behavioral experiments in an inverse reinforcement-learning framework \citep{schultheis2021inverse,ashwood2022dynamic},  our approach is the first to integrate into this endeavor an assessment of deviations from what the experimenter believes to be {{ the optimal behavior}.
Designs for fixed interval experiment {vary widely.} We show how our method is useful for the standard fixed interval design, but it may apply equally well to other variations of these studies. Further, the idea of learning a reward that is a divergence from optimality can be quite general, and may apply to other  more complex experiments,  such as intermittent reinforcement designs.  


\section{Existing approach}
Let the random variable $S$ denote the mouse's exposure, {where} $S=1$ if the mouse was exposed to a {toxicant} and 0 otherwise. We also introduce the random variable $A$ for the mouse's action, which could be, for example, the number of presses the animal makes per session {in the experiment shown in Figure \ref{fig:fixedIntervalExperiment}} (hence $A$ might be a summary statistic of the animal's actions over time). 
Some true model, $\pi_0(A=a|S=s)$ ({where the subscript $0$ indicates} that it is the true model), governs the animal's actions given their exposures,  which is, in a reinforcement learning framework, the mouse's true ``behavioral policy" \citep{sutton2018reinforcement}.
It is common in these types of fixed interval experiments to specify a model, $\pi_b$ ({where the subscript} $b$ indicates that it is a specified behavioral model), for the true behavioral policy, $\pi_{0},$ as, for example,
\begin{equation}
\label{eq:BehaviorPolicyLM}
\pi_0(A=a|S=s) = N(b_0 + b_1^T S,\sigma^2),
\end{equation}
which is not unreasonable for certain actions. Note that the model in (\ref{eq:BehaviorPolicyLM}) can arise in indirect ways; even a reasonable procedure, such as comparing averages over groups, is equivalent to assuming (\ref{eq:BehaviorPolicyLM}) as a model of the mouse’s policy.
The model in (\ref{eq:BehaviorPolicyLM}) is equivalent to an Analysis of Variance (ANOVA) \citep{scheffe1999analysis}, and indeed the ANOVA approach is commonly used for the analysis of fixed interval experiments. 
However, specification of a behavioral policy $\pi_b$ that matches the true density, $\pi_{0},$ is quite difficult in practice.   The true conditional density $\pi_0$ could be, in general, a highly complex function, and the space of possible choices for a model, $\pi_b,$ for $\pi_0$ is vast. 
The model for the behavioral policy, $\pi_b,$ is therefore likely to be misspecified. 
 For example, even though it is a saturated model, a linear ANOVA model breaks down if the action is nonnegative, as is the case in many fixed interval experiments, since the number of presses in the session is the response variable. A log linear model might be more appropriate, but so might a Gamma distribution, for example.  It is difficult to know which one to choose.  In the simulations, we will show a case in which the true generating behavioral distribution is a complex Gamma distribution, and therefore it is quite different from a linear ANOVA, a log linear model, or anything that is possible to correctly specify.

Note that although, as mentioned, it is difficult to know whether an ANOVA is the correct model, using an ANOVA (or just comparing average behaviors over groups) is a reasonable choice for this type of experiment. Further, scientists often use a battery of measures to characterize a mouse's behavior, and ANOVA is just one component of this.  However, in this work, we seek to give an alternative to the  ANOVA, which can possibly be used in conjunction with an ANOVA and other measures, enriching the overall analysis of the experimenter.

\section{Methodology: optimality based-reward learning}

Because of the difficulty in specifying the {model for the true behavioral} policy,  $\pi_0,$ we choose to avoid modeling it altogether. We instead model the mouse's reward function.  A reward function, like the behavioral policy, {can} be difficult to specify. However, we sidestep the need to specify the reward function by incorporating knowledge of the optimal behavior, which is known to the experimenter.  We parameterize the reward, and, in learning this parameter, obtain a novel summary of the alteration caused by the {toxicant}.

\subsection{Reward as divergence from optimality}
 Modeling a reward function is often more difficult than modeling a behavioral policy. However, we include in our reward function information known to the experimenter about optimality, which considerably simplifies the task.  
In other words, in a typical fixed interval experiment, such as the one shown in Figure \ref{fig:fixedIntervalExperiment},  we can establish the optimal action, $A^*,$  based on our understanding of the experiment.  For example, in the experiment mentioned in Figure \ref{fig:fixedIntervalExperiment}, 
one might expect near-optimal mice to allocate responses in a scallop pattern, {in which they rarely press the lever right after receiving a food pellet, and then slowly increase their responses as the 60 s mark, which ends the fixed interval and portends a new food pellet, approaches \citep{skinner1938behavior}}. 
 For $X\in\mathbb{R}^K,$ define $||X||_2^2={\sum_{i=1}^KX_i^2}.$ Let $A^*$ denote the optimal action. For a particular mouse taking action $A$, we can define a reward that penalizes divergence from optimality as
\begin{equation}
\label{eq:objReward}
R(A)=-||A-A^*||_2^2.
\end{equation}
We negate the divergence, so a mouse will achieve the highest reward if it takes the optimal action, $A^*.$  %
Note that this is an \textit{objective} reward.  In other words, if mouse 1 takes action $A_1$, which is 1 unit away from the optimal action, and mouse 2 takes action $A_2,$ which is 2 units away from the optimal action, then, according to (\ref{eq:objReward}),  mouse 1 has a higher objective reward than mouse 2; i.e., $R(A_1)> R(A_2)$.  

\subsection{A parameterized, subjective reward}
Now, let us consider a definition of a personal, \textit{subjective}  reward, in which we will introduce a parameter, $\theta$, which is specific to a particular type of mouse (e.g., mice that are exposed to a {toxicant}). Concretely, a subjective reward can be written, for  $\theta\in [0,1]$ specific to a mouse that takes action $A$, as
{
\begin{equation}
\label{eq:subReward}
R(\theta,A) = -||(A-A^*)||_2^2\theta.
\end{equation}
}
 Now, depending on $\theta,$ this reward may not correspond to the objective reward in (\ref{eq:objReward}).  Suppose again we have mouse 1 who is 1 unit away from the optimal action, and mouse 2 who is 2 units away from the optimal action.
 Suppose that mouse 1 belongs to a group with $\theta_{1}=1$ and mouse 2 belongs to a group with {$\theta_{2}=0.25.$} Then these two mice have equal subjective rewards, $R(\theta_{1},A_1) =R(\theta_{2},A_2),$ even though, as we saw before, their objective rewards differ: $R(A_1)>R(A_2)$. In general, mice that diverge from the optimal action have a low reward. If a mouse never gets tired and also receives no benefit from the food pellet, it might have $\theta=0;$ which implies that its action does not affect its reward. 

\subsection{Equality of subjective rewards}

We assume that 
every animal in a fixed interval experiment takes actions that maximize their personal, subjective reward, as defined in Equation (\ref{eq:subReward}). 
{ We make the following remark, which is essential to performing estimation.} 

\begin{remark}
\label{rem:equalSub}
{Any two animals will have equal subjective rewards, since the animal with the lower reward can change their actions if this is not the case.}

\end{remark}
 One can see the implications of Remark \ref{rem:equalSub} by examining the reward function. 
In particular, for mouse $i$ from the exposed group E (with parameter $\theta_E$) taking action $A_i$  and for mouse $j$ from the control group C (with parameter $\theta_C$) taking action $A_j$, we have that $R(\theta_E,A_i)=R(\theta_C,A_j).$   This only occurs if, using the definition of $R$, {$-||(A_i-A^*)||^2_2\theta_E=-||(A_j-A^*)||^2_2\theta_C.$}  Therefore, we are stating in Remark \ref{rem:equalSub} that if mouse $i$ diverges more from the optimal action than mouse $j$, then mouse $i$ tolerates divergence from optimality more than mouse $j$, or that $\theta_E<\theta_
C$.

\subsection{Identifying exposure-specific parameters}

Now, let us create two groups corresponding to the exposed, $E,$ and control, $C,$ mice, indexed by $\theta_E$ or $\theta_C,$ respectively. The parameter $\theta_{0,E},$ the exposed mouse's true tolerance for divergence from optimality, or the alteration caused by the {toxicant}, is the estimand of interest in this work. { The closer $\theta_{0,E}$ is to zero, the more the exposed animals tolerate a divergence from optimality relative to the controls.}
{ As discussed in Appendix \ref{app:identifiability}, the parameter $\theta_E$ is not identifiable (if we were to construct a system of equations using the form of $R$ in (\ref{eq:subReward}), the solution for $\theta_E$ will always be in terms of $\theta_C$)}. Let us impose the following constraint, which identifies $\theta_E$,
\begin{equation}
\label{eq:identifiability}
\theta_E+\theta_C=1.
\end{equation}
The constraint (\ref{eq:identifiability}) allows us to rewrite the subjective reward in (\ref{eq:subReward}) as a function of the exposure, $S,$
{
\begin{align}
\label{eq:subjectiveRdependsonS}
\notag R(\theta_E,\theta_C,A,S)&={-||(A-A^*)||_2^2(\theta_E S + \theta_C(1-S))}\\
\notag &=-||(A-A^*)||_2^2(\theta_E S + (1-\theta_E)(1-S))\\
&=:R(\theta_E,A,S).
\end{align}
}
Under (\ref{eq:subjectiveRdependsonS}), a mouse who was exposed has $S=1,$ and, therefore, $\theta_E,$ and a mouse who is a control has $S=0$ and, therefore, $1-\theta_E=\theta_C.$
\subsection{Interpreting the estimand}
\label{sec:interpretThetan}

Under the constraint in (\ref{eq:identifiability}), any tolerance of divergence in the exposed group is associated with a corresponding lack thereof in the control group.  Note that $\theta_E=\theta_C$ only if $\theta_E=\theta_C=0.5,$ in which case both groups tolerate a divergence from optimality equally.  If $\theta_E<0.5,$ then the exposed mouse tolerates divergence from optimality more than the control mouse; i.e., the exposure is a {toxicant}.  If $\theta_E>0.5,$ then the exposed mouse tolerates divergence from optimality less than the control mouse; i.e., the exposure is beneficial (e.g., a medicine or essential nutrient).  

\subsection{Estimation}

Now that we have, in (\ref{eq:subjectiveRdependsonS}), a parameterized model for the subjective, exposure-specific reward of any mouse, we can use it to find an estimator, $\theta_{n,E},$ for $\theta_{0,E}.$ Note that we subscript our estimands by 0 and our estimators by $n.$   To estimate $\theta_{0,E}$, we rely on Remark \ref{rem:equalSub}, which states that any two animals { will  have the same subjective reward. 
We assume that any difference between subjective rewards in the observed animals { whether they are exposed or not} is, therefore, due to noise. 

We can construct an objective function that represents the difference between subjective rewards for all pairs of mice in the experiment.} Concretely, if we have observed $n$ independent and identically distributed state-action pairs, $(A_1,S_1),\dots,(A_n,S_n),$ each corresponding to one mouse in a fixed interval experiment, we can compare each mouse  with all the other mice,
\begin{equation}
\label{eq:pwdiff.Objective}
\Psi_n(\theta_E) = \frac{1}{n^2}\sum_{i=1}^n\sum_{j=1}^n\left(R(\theta_E,A_i,S_i)-R(\theta_E,A_j,S_j)\right)^2.
\end{equation}
{Note that each mouse is compared to all other mice, regardless of group membership, but, within a group, the mice are constrained to have equal $\theta$, and, hence, the out of group comparisons drive the differences in $\theta_{E,n}$ during estimation.} Note that, based on (\ref{eq:subjectiveRdependsonS}), when $S=1,$ $R(\theta_E,A_i,S_i)$ depends on $\theta_E,$ whereas when $S=0,$ $R(\theta_E,A_i,S_i)$ depends on $1-\theta_E=\theta_C,$ but we only have to estimate $\theta_E$ because of the constraint in (\ref{eq:identifiability}).  
 { One can find $\theta$ that makes these pairwise differences as  small as possible, which aligns with Remark \ref{rem:equalSub}.  Doing so drives $\theta$ to be small (or the tolerance for divergence to be large) for mice that act suboptimally, and vice versa for mice that act optimally. 
 We then have an estimator for $\theta_{E,0},$}
\begin{equation}
\label{eq:pwdiff}
\theta_{E,n} = \arg\min_{\theta_E}\Psi_n(\theta_E).
\end{equation}

\section{Theory}
We will now discuss some of the theoretical properties of our objective function in (\ref{eq:pwdiff.Objective}) and our estimator, $\theta_{E,n}.$
We appreciated personal correspondence with \citet{whuber}, which keyed us in to the following remark.

\begin{remark}
\label{rem:pwdiff2var}
We have that the objective function in Equation (\ref{eq:pwdiff.Objective}) is equal to two times the sampling variance, $\text{var}_n(R);$ i.e.,
$\Psi_n = 2\text{var}_n(R).$ 
\end{remark}
For a proof, see Appendix \ref{app:proof:rem:pwdiff2var}. { Hence, finding a parameter  $\theta$ that makes the subjective rewards of any two observed mice as close as possible is equivalent to finding a parameter $\theta$ that minimizes the empirical variance of the subjective rewards.}
 Remark \ref{rem:pwdiff2var} is interesting in its own right, and it will also help us in the theory that follows.  First, we will consider consistency of the proposed objective function.

\begin{theorem}
\label{thm:consistPairwiseObj}
The estimator $\Psi_n$ is consistent for $\Psi_0$, where $\Psi_0= \text{var}_0(R(\theta_E,A_1,S_1)),$ and $\text{var}_0$ is the population variance.
\end{theorem}

For a proof, see Appendix \ref{app:proof:thm:consistPairwiseObj}.

\noindent One can also examine the rate {of convergence of} the objective function.


\begin{theorem}
\label{thm:rate}
    We have that \[\sqrt{n}\frac{1}{n^2}\sum_{i=1}^n\sum_{j=1}^n\left(R(\theta_E,A_i,S_i)-R(\theta_E,A_j,S_j)\right)^2=O_P(1).\]
\end{theorem}

For a proof, see Appendix \ref{app:proof:thm:rate}.

\noindent We finally establish consistency of $\theta_{E,n}$ in the following theorem.

\begin{theorem}
\label{thm:consistencyThetaEn}
    Assume that $A$ in practice only attains observed realizations, $a,$ that are bounded; i.e., $|a|<\infty$ for all $a$. We then have consistency of $\theta_{E,n},$ or that \[\theta_{E,n}\gop \theta_{E,0}.\]
\end{theorem}
For a proof, see Appendix \ref{app:proof:thm:consistencyThetaEn}.


\noindent Since our estimator is a finite-dimensional minimum of an objective function, we have obtained consistency fairly easily.  Note as well that this applies to arbitrarily complex actions; as long as its divergence from the optimal action is bounded, $A$ can be a vector (e.g., for time-dependent experiments), a matrix (e.g., for time-dependent experiments with grouped experimental units), or even a tensor.

\section{Simulations}
\label{sect:sim}

Suppose the true behavioral policy, $\pi_0,$ that governs the animal's actions $A$, as a function of the exposure, $S,$ is a distribution {that is difficult to specify,} such as a Gamma distribution ($\Gamma$) whose shape $\alpha$ depends on $S$ in a complex way. For example, suppose the true, data-generating policy is
\begin{equation}
\label{eq:simDis}
\pi_0(A=a|S=s)=\Gamma\left(\alpha=(2\epsilon_1)^s (1\epsilon_2)^{1-s},\beta=1\right),
\end{equation}
where $\epsilon_1,\sim N(\mu_1=2,\sigma^2_1=1)$ and $\epsilon_2\sim N(\mu_2=2,\sigma^2_2=4).$
This model leads to larger $A,$ or more activity, when $S=1$ on average; {in summary, this model states that exposure causes hyperactivity in a complex way.}
Note that the true behavioral policy in (\ref{eq:simDis}) is unknown to the experimenter.  The experimenter may model the behavioral policy nevertheless using (\ref{eq:BehaviorPolicyLM}), as is commonly done, or they may use the proposed method by optimizing  (\ref{eq:pwdiff}). Set the sample size to be $n=50$ (this is small, which is reflective of the sample size in true fixed interval mouse experiments) and the number of Monte-Carlo datasets to be $M=2000$.    We  compare results when one  fits a linear model to this data (i.e., (\ref{eq:BehaviorPolicyLM})) to obtain an estimate of the exposure effect, $b_{1,n},$ to results when one uses the proposed method to optimize (\ref{eq:pwdiff}) for $\theta_{E,n}$, shown in Figure \ref{sim}. The exposure assignment, $S$, is drawn randomly with probability 0.5. 
\begin{figure}[htp!]
    \centering
    \includegraphics[width=0.7\textwidth]{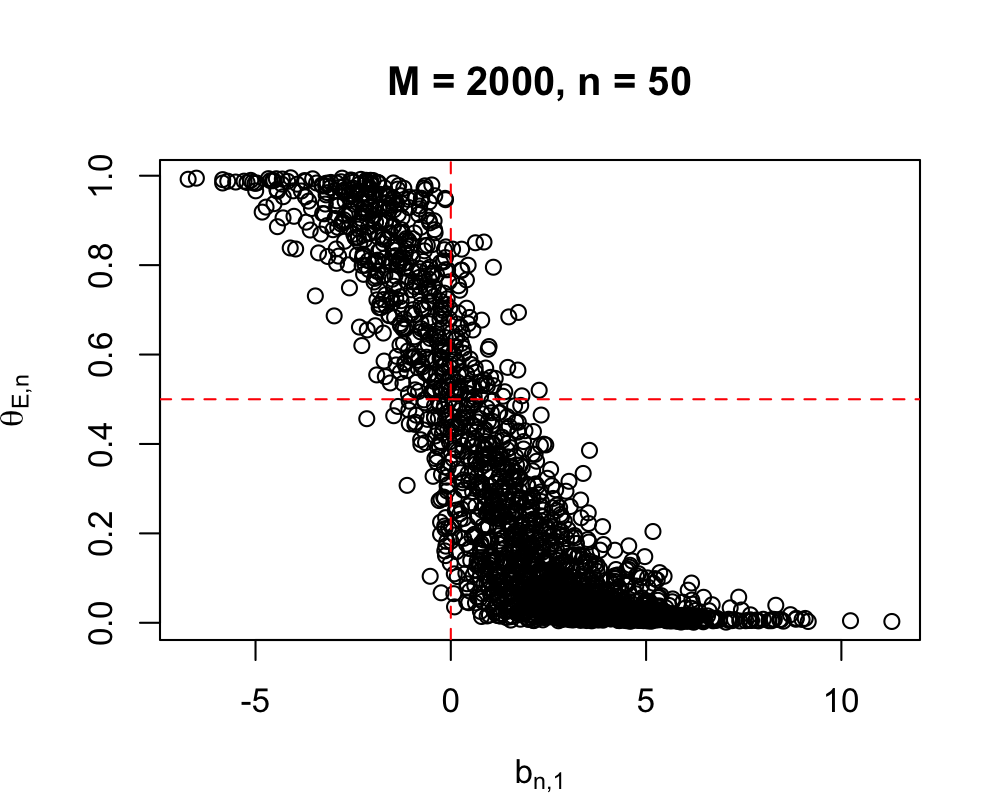}

    \caption[Optimality-based reward learning simulation]{Estimates of $\theta_{E,n}$ and $b_{1,n}$ over Monte-Carlo datasets. { Red dashed lines correspond to $\theta_{E,n}=0.5$ and $b_{n,1}=0$}.}
    \label{sim}
\end{figure}


This is a very small sample of noisy data, but we see {in Figure \ref{sim}}  that $b_{1,n}$ is often greater than zero, indicating that the exposure increases activity. This is accurate based on the model $\pi_0$ (the model $\pi_0$  in (\ref{eq:simDis}) leads to higher activity when $S=1$ on average). However, the parameter $b_{1,0}$ in the misspecified model  does not correspond to the target estimands in the true model, $\pi_0,$ in (\ref{eq:simDis}), which  are $\alpha$, $\beta$, $\mu_1,\mu_2,\sigma_1,\sigma_2$.  Yet, our estimate of $\theta_{E,n}$ which, when less than $0.5$ indicates that the exposure increases activity (by the discussion in Section \ref{sec:interpretThetan}), usually agrees ({as shown in Figure \ref{sim}}) with $b_{1,n}$ in that $\theta_{E,n}$ 
is usually less than 0.5, also indicating that exposure increases activity. Although $b_{1,n}$ and $\theta_{E,n}$  usually, in a broad sense, come to the same conclusion  about whether the exposure harms or benefits the mice ($\theta_{1,n}<0.5$ for 73.85\% of the simulated datasets and $b_{1,n}>0$ for 74.15\% of the simulated datasets), the proposed method makes no assumption about the form of the behavioral policy in (\ref{eq:simDis}), and thus, in general, it is more robust.  Further, the magnitude of $\theta_{E,n}$ and $b_{1,n}$ have different interpretations.
Unlike $b_{1,n},$ $\theta_{E,n}$ does correspond to a true estimand $\theta_{E,0},$ and $\theta_{E,0}$ does have a valid interpretation as the  tolerance for diverging from optimality in the exposed mice. 
The magnitude of $\theta_{E,n}$ is meaningful, whereas the magnitude of $b_{1,n}$ is only meaningful under correct specification of (\ref{eq:BehaviorPolicyLM}).  Note that if we knew that the true model for $\pi_{0}$ was (\ref{eq:BehaviorPolicyLM}), it would be better to estimate $b_{1,0}.$  However, we would not know this in practice.

In conclusion,  we see that the estimate $b_{1,n}$ does not correspond to any estimand in $\pi_0$ (equation (\ref{eq:simDis})), whereas the estimate $\theta_{E,n}$ corresponds to $\theta_{E,0},$ the exposed group's tolerance for divergence from optimality, regardless of the true behavioral model $\pi_0$.  


{
\section{Real Data analysis}
We analyze data from a fixed interval experiment in which mice are exposed to ambient iron \citep{eckard2023neonatal}, which acts as a cerebral neuro{toxicant}. The fixed interval length was 60 s.  Since the session is of length 30 minutes, a mouse can obtain up to  30 food pellets in one session. The sessions are performed $K=25$ times (and the mice tend to improve with each session).  { We show histograms for the number of presses in each 5 second time bin (over all sessions) in Figure \ref{fig:fi.data}. We observe that the iron-exposed mice are generally more hyperactive than the air-exposed (control) mice, and that both groups show an increase in their responses over time (a scallop pattern).}

\begin{figure}
    \centering
    \includegraphics[width=0.8\textwidth]{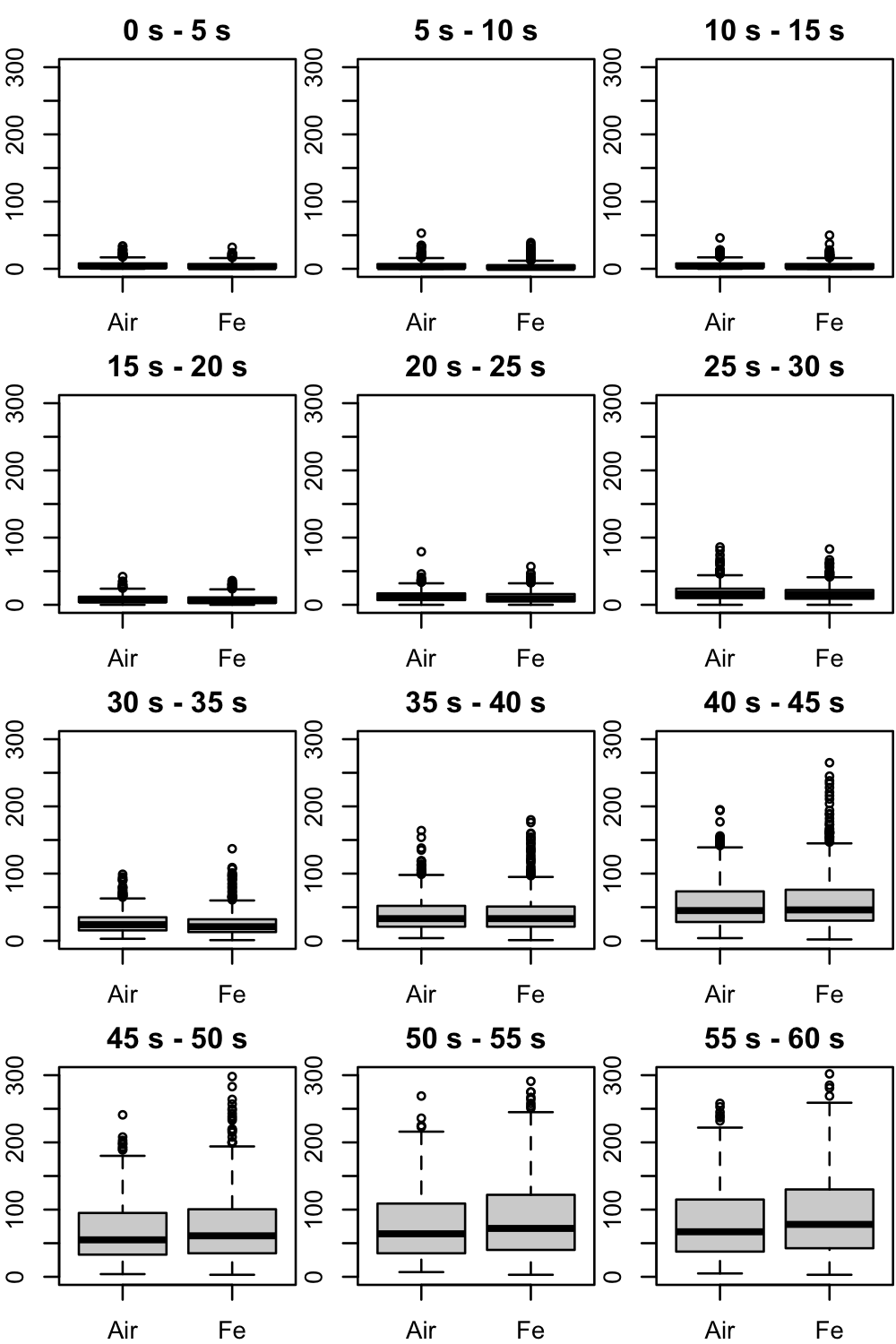}
    \caption{Histograms for the number of presses in each 5 second time bin over all sessions. Data is from \citet{eckard2023neonatal}.}
    \label{fig:fi.data}
\end{figure}

First, we describe how we define the actions of the mice.  In the data from \citet{eckard2023neonatal}, we observed, for mouse $i$ within a session $k,$ a vector { of length 12}, \[A_{i,k}=(N_{i,k,0-5},N_{i,k,5-10},N_{i,k,10-15},\dots,N_{i,k,50-55},N_{i,k,55-60})^T,\] where $N_{i,k,t_1-t_2}$ is the number of lever presses that occurred from $t_1$ seconds to $t_2$ seconds by mouse $i$ during session $k$.  We average over the  $K=25$ sessions.  Hence, we define  the action for mouse $i,$ averaged over sessions, as \[A_{i}=\frac{1}{K}\sum_k(N_{i,k,0-5},N_{i,k,5-10},N_{i,k,10-15},\dots,N_{i,k,50-55},N_{i,k,55-60})^T.\] 
Note that the action is a vector instead of scalar, as in the simulations.  The flexibility with which one can represent the action is a strength of the proposed method.
 Conversely, if one models the same data using the approach in (\ref{eq:BehaviorPolicyLM}), one must first summarize  (assuming one does not wish to employ complex, time-varying response models) the vector $A_i$, with component $k$ defined as $(A_i)_k,$ as $\tilde{A}_i = \frac{1}{12}\sum_{k=1}^{12}(A_i)_k$, and then model 
\begin{equation}
\label{eq:real.data.lm}
    \pi_0(\tilde{A}_i=a|S_i=s)= N(b_0 + b_1 s,\sigma^2).
\end{equation}

 Let the bin midpoint times be defined as \[M=(2.5, 7.5,\dots,57.5)^T.\] Let the optimal action, $A^*=(1,0,\dots,0),$ {a vector of length 12,} be the strategy in which one makes only one press during the 0-5 second interval. Note that the initial lever press in a session starts a 60 second refractory interval; any press during this refractory interval does not yield a food pellet. A press is followed by receipt of a food pellet only if it occurs after this 60 second refractory interval has ended. The lever press that results in a food pellet initiates a new 60 session interval with the same constraints \citep{eckard2023neonatal}. 
Hence, the optimal action, $A^*$, is to press in the $N_{0-5}$ bin. If a mouse instead presses later in the interval, the pauses can add up so that a mouse might not receive the maximum number of food pellets in a session, which lasts exactly 30 minutes.

Although $A^*$ is optimal, it is not attainable by a mouse unless the mouse can be sure when the 0-5 second interval occurs (i.e., if the mouse has a stopwatch), so we do not measure the divergence from $A^*$, but from a scaled version of it; i.e., we define the reward for mouse $i$ as 
{
\[R(\theta_{E},A_i,S_i)=-||(A_i-A^*)\odot(60-M)||_2^2(\theta_{E}S_i+(1-\theta_E)(1-S_i)).\] 
}
This reward forgives the first press and promotes a non-hyperactive scallop pattern (for more on the scallop, see e.g.  \citet{dews1978studies,skinner1938behavior}), which would be an optimal action for a mouse without a stopwatch or others means of knowing the time exactly.

 Given our definition for $R,$ we finally optimize (\ref{eq:pwdiff}) to obtain estimate $\theta_{E,n}=0.435<0.500,$ which implies that ambient iron alters the reward function of the mice, promoting suboptimal behavior.  Conversely, using the ANOVA specified in (\ref{eq:real.data.lm}), one obtains $b_{1,n}=4.16.$ As discussed,  if the true behavioral model is not correctly specified in (\ref{eq:real.data.lm}), $b_{1,n}$,  unlike $\theta_{E,n}$,   will not correspond to a true estimand. 
   {In Figure \ref{fig:Rviz}, we include a visualization of the average reward (defined in (\ref{eq:subjectiveRdependsonS})) for the exposed and control mice as a function of $\theta$. Note that the average rewards are approximately equal at $\theta_{E,n}$  (note that (\ref{eq:pwdiff.Objective}) requires equality for the rewards of each animal rather than equality of the averages of the rewards, so $\theta_{E,n}$ is not exactly where the curves cross, but it is very close). We see that $\theta_{E,n}<0.5.$  As discussed in Section \ref{sec:interpretThetan}, this implies that the exposed mice tolerate divergence from optimality more than  the control mice.}  
 \begin{figure}
     \centering
     \includegraphics[width=\textwidth]{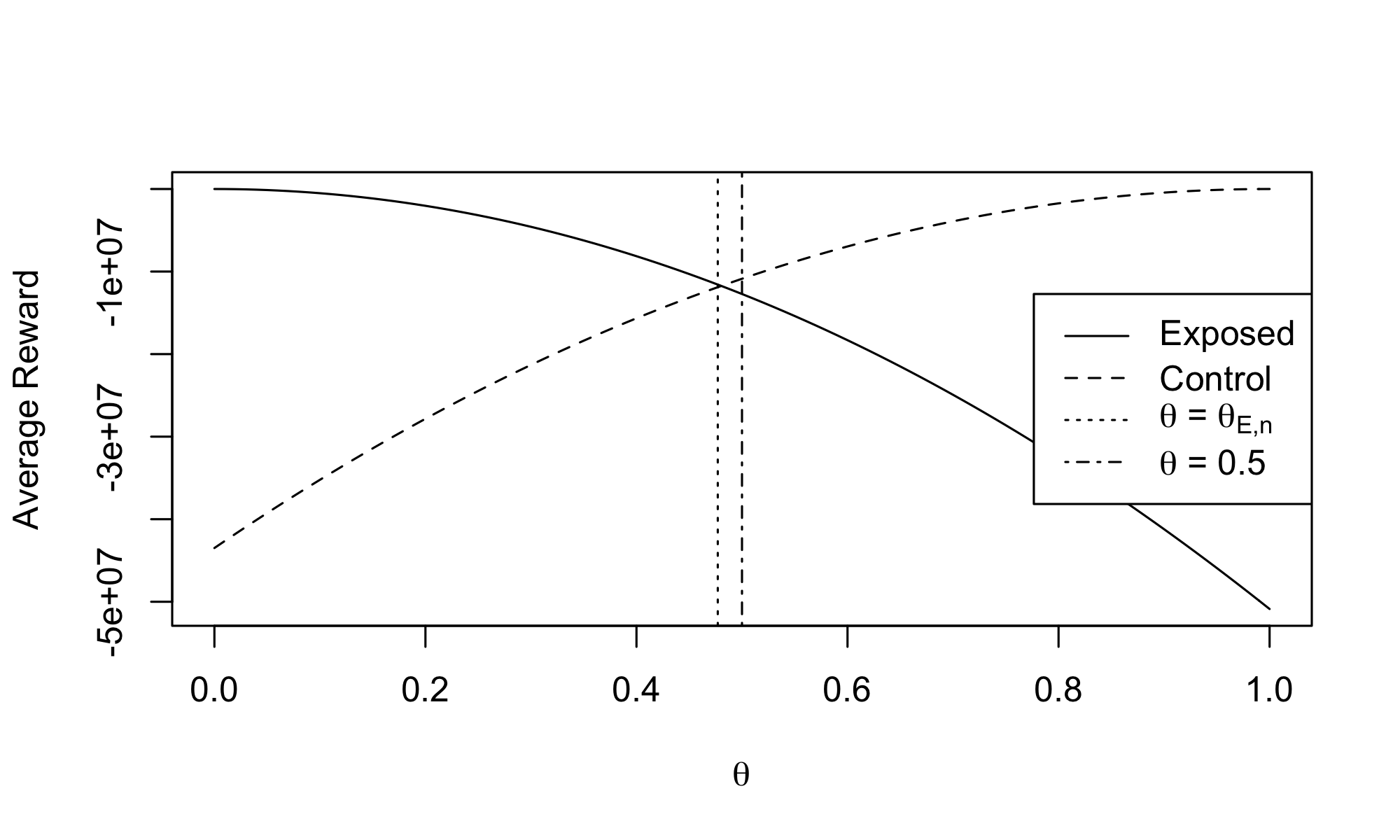}
     \caption{{The average rewards for the mice in the exposed and control groups are shown as a function of $\theta,$ the tolerance for divergence from optimality. }}
     \label{fig:Rviz}
 \end{figure}

}

\section{Discussion}

We have proposed a method for analyzing a fixed interval experiment that  sidesteps the difficulties associated with specifying a behavioral policy, or a conditional density of actions given exposures, {by instead modeling the animal's reward function. The proposed method also sidesteps the need to specify a reward function explicitly by incorporating knowledge known to the experimenter on optimality.  The parameters of the reward function are finite-dimensional, but make very few assumptions about the data generation itself.}  Ultimately, better analysis of fixed interval experiments allows us to better utilize resources when designing these experiments  and potentially to better understand the effects of {toxicant}s that are present in our environment, which is a necessary step toward protecting humans from exposure.

We {first} discuss some limitations and future directions of the real data analysis.  In this study, we analyze the animal behavior averaged over all the sessions. However, the “scallop” behavioral pattern should improve for each animal as the sessions progress. Future analysis might consider computing the estimator as it changes over sessions{, which would correspond to the animal's behavior approaching the behavioral optimality over time, which could then further be assessed as it differs between the control and exposed mice}. Note also that the bins are 5 seconds long  in this study. Changing the bin size might change results. Ideally, if one knew the time of each press, one could adapt the proposed method to penalize the deviation of each press from optimality, rather than averaging over all presses within a given bin and computing the deviation of this average from optimality, as we have done here. Finally, the data is collected in a free operant system; i.e., the animals can skip intervals early on in training. To better quantify the scallop, one could use a system with discrete operant trials; i.e., one could give animals an inter-trial interval time, where the lights go off to signal that the 60 second interval is over and then go back on to start another interval. The skipping of intervals creates noise in the analysis. The skipping is mostly a problem during early sessions, before the animals have adequately learned the task. One could potentially assess the degree to which animals skip reinforcers in the early sessions, and one could eliminate early sessions from the analysis, if needed. Alternatively, one could use an $L_1$ norm in (\ref{eq:subReward}), which is more robust to noise created by skipping.  {Another interesting direction might be to develop a version of the metric that conditions on more information, in the form of covariates. The estimator we provide marginalizes over these covariates, but a conditional model might be of scientific interest to provide a more subject-specific analysis.}

We now discuss some limitations of the proposed method. When defining the reward under our proposed method, we do have to choose a metric, such as an $L_1$ or $L_2$ norm, for the distance between the observed and optimal actions. The choice of norm, however, is not an assumption in the way that a specific model for $\pi_0$ is an assumption.  In other words, if we choose an $L_2$ norm, our conclusions are that the tolerance for $L_2$ norm divergence from the optimal is estimated by $\theta_{E,n},$ but if we choose the $L_1$ norm, our conclusion is that the tolerance for $L_1$ norm divergence from the optimal is $\theta_{E,n};$ both of these can be true statements, {since the magnitude of $\theta$ does not have any intrinsic meaning}.  However, if we misspecify a model $\pi_b$ for $\pi_0,$ a statement that $b_{n,1}$ corresponds to some estimand in $\pi_0,$ which is what underlies the assumption that $\pi_b$ is our chosen model, is false. 
  Note also that in the proposed approach we are choosing to focus only on divergence from optimality rather than on the direction of divergence. By providing a sign on the coefficient, an ANOVA does provide a sort of inference for the sign of the effect of the {toxicant} (whether the {toxicant} increases of decreases actions).  However, the correctness of the ANOVA coefficient sign is dependent upon the correct specification of the ANOVA. One could glean a rough estimate of the sign of the effect in our case by taking a contrast of the exposed and control group average actions. In general, however, in many toxicological experiments, it may be useful to establish whether there is an effect at all at a certain exposure level and dose, a fact that can be then augmented by an understanding of the biological mechanism of the effect to determine the likely direction of the effect and implications for those who are exposed. 


One additional contribution of our work here is to fully cast the fixed interval mouse experiment within a reinforcement learning framework. When viewing a fixed interval experiment in a reinforcement learning framework, other possible statistical summaries besides the proposed $\theta_{E,n}$ also come to mind. Some might be quite useful, and they are discussed in terms of their relationship with $\theta_{E,n}$ in Appendix \ref{app:otherStatistics}. However,  $\theta_{E,n}$ has certain favorable properties. For example, the magnitude of  $\theta_{E,n}$ is bounded in the range $[0,1]$ regardless of the scale of $A,$ which cannot be said for other more ad-hoc statistics.  Further, $\theta_{E,n}$ is the solution to an optimization problem, and, besides giving the statistic $\theta_{E,n}$ concrete meaning ({although the magnitude of the statistic does not have inherent meaning, it can be interpreted as a tolerance for divergence, which is higher when it is near zero}), this optimization allows one can to use results such as those shown in Theorems \ref{thm:consistPairwiseObj} and \ref{thm:rate} to ultimately draw conclusions on the consistency of $\theta_{E,n}$ in Theorem \ref{thm:consistencyThetaEn}. Further, although the statistic $\theta_{E,n}$ requires very few assumptions, it is still parametric, and therefore will have favorable convergence properties.  In general, with other methods, it is difficult to obtain the combination of a small number of assumptions, the benefits of a parametric model, and the underlying optimization framework that are jointly offered by $\theta_{E,n}$. It is difficult to interpret behavior from any one summary variable, so scientists use behavioral batteries in multiple domains, and look for broad patterns. We hope that the proposed statistic provides a new addition to these batteries.




\section{Acknowledgements}

%
%
%
%
%
%
%

This work is solely the responsibility of the authors and not the funding agencies. Research  reported  in  this  publication  was  supported  by  the  National  Institutes  of  Health (NIH) under award numbers    T32ES007271 (National Institute of Environmental Health Sciences (NIEHS)) and 
 T32GM007356 (National Institute of General Medical Sciences (NIGMS)).

\appendix
 \renewcommand\thefigure{\thesection.\arabic{figure}}  
 \setcounter{figure}{0}   
 
\section{Appendix}
\subsection{Identifiability}
\label{app:identifiability}
Consider the population exposed mouse and population control mouse, who take actions $a_E,a_C.$  We are claiming that their subjective rewards are equal, or that 
\begin{equation}
\label{eq:equalSubR}
R(\theta_C,a_C)=R(\theta_E,a_E).
\end{equation}
We will now give a small example showing non-identifiability without the constraint in (\ref{eq:identifiability}), which was $\theta_E + \theta_C =1$.  Suppose $a_E=3, a_C=2, A^*=1.$ We will start with (\ref{eq:equalSubR}) and show that, under these conditions, we have more than one solution for $\theta_E.$ Assume no constraint on $\theta_E$ and $\theta_C$ and write, using the definition of $R$ in (\ref{eq:subReward}),
{
\begin{align}
\label{eq:nonIdentSoln}
R(\theta_C,a_C)=R(\theta_E,a_E) 
\notag &\iff -((a_c-A^*)^2\theta_C) = -((a_e-A^*)^2\theta_E)\\
\notag &  \iff -((2-1)^2\theta_C)=-((3-1)^2(\theta_E))\\
\notag & \iff \frac{\theta_C}{\theta_E} = 4\\
& \iff \theta_E=\theta_C/4. 
\end{align}
Hence, for $\theta_C=1$ we  obtain $\theta_E=\frac{1}{4}$ and for $\theta_C=2$ we obtain $\theta_E=1/2,$ and both imply (\ref{eq:equalSubR}), implying a lack of identifiability.
When we impose identifiability constraint $\theta_E+\theta_C=1$ from (\ref{eq:identifiability}). Starting from (\ref{eq:nonIdentSoln}), we have uniqueness of $\theta_E$
\begin{align*}\theta_E &= \theta_C/4\\ &\iff \theta_E = (1-\theta_E)/4\\ &\implies \theta_E = 1/5<1/2.\end{align*}
}
\subsection{Proof of Remark \ref{rem:pwdiff2var}}
\label{app:proof:rem:pwdiff2var}
\begin{proof}
    Note, letting $R_i=R(\theta_E,A_i,S_i)$ for compactness, that our objective in (\ref{eq:pwdiff.Objective}) is
\begin{align}
\label{eq:breakdown}
   \notag \Psi_n &=  \frac{1}{n^2}\sum_{i=1}^n\sum_{j=1}^n\left(R_i-R_j\right)^2\\
   \notag &=\frac{1}{n^2}\sum_{i=1}^n\sum_{j=1}^n\left(R_i-ER_1+ER_1-R_j\right)^2\\
   \notag    &=\frac{1}{n^2}\sum_{i=1}^n\sum_{j=1}^n\left((R_i-ER_1)-(R_j-ER_1)\right)^2\\
   \notag &= \frac{1}{n}\sum_i (R_i-ER_1)^2 
     +\frac{1}{n}\sum_j (R_j-ER_1)^2\\
     \notag & \ \ \ \ \ -2 \frac{1}{n^2}\sum_i\sum_j (R_i-ER_1) (R_j-ER_1)\\
     \notag  &= \frac{1}{n}\sum_i (R_i-ER_1)^2 
    +\frac{1}{n}\sum_j (R_j-ER_1)^2\\
     & \ \ \ \ \ +2 \frac{1}{n}\sum_i (R_i-ER_1)\frac{1}{n}\sum_j(R_j-ER_1)\\
   \notag &= 2\text{var}_n(R_1),      
\end{align}
\end{proof}

\subsection{Proof of Theorem \ref{thm:consistPairwiseObj}}
\label{app:proof:thm:consistPairwiseObj}
\begin{proof}
Note that by Remark \ref{rem:pwdiff2var}, the objective function in (\ref{eq:pwdiff}) is equal to $2\text{var}_n(R),$
where $\text{var}_n(R)$ is the sampling variance. Note that by the Law of Large Numbers, $2\text{var}_n(R_1) \gop 2 \text{var}_0(R_1),$ where $\text{var}_0$ is the population variance.  
\end{proof}

\subsection{Proof of Theorem \ref{thm:rate}}
\label{app:proof:thm:rate}
\begin{proof}
    Start with  (\ref{eq:breakdown}) derived in the proof of Remark \ref{rem:pwdiff2var}, or that 
    \begin{align*}
  \Psi_n&= \frac{1}{n}\sum_i (R_i-ER_1)^2 
    +\frac{1}{n}\sum_j (R_j-ER_1)^2\\
    &\ \ \ \ +2 \frac{1}{n}\sum_i (R_i-ER_1)\frac{1}{n}\sum_j(R_j-ER_1).   
\end{align*}
    Multiply through by $\sqrt{n}$ to obtain 
        \begin{align*}
   \notag \sqrt{n}\Psi_n
        &= \frac{\sqrt{n}}{n}\sum_i (R_i-ER_1)^2 
    +\frac{\sqrt{n}}{n}\sum_j (R_j-ER_1)^2\\
    &\ \ \ \ \ + 2\frac{\sqrt{n}}{n}\sum_i (R_i-ER_1)\frac{1}{n}\sum_j(R_j-ER_1)\\
    &= (I) + (II) + (III).
\end{align*}
Note that $(I)=(II),$ and both are scaled averages of iid variables and, therefore, converge in distribution to normal random variables by the Central Limit Theorem. Hence, $(I)$ and $(II)$ are $O_p(1).$
The term $(III),$ the product of a normally distributed random variable and an average that converges to zero, disappears by Slutsky's theorem.
\end{proof}

\subsection{Proof of Theorem \ref{thm:consistencyThetaEn}}
\label{app:proof:thm:consistencyThetaEn}
\begin{proof}
By the algebra shown in the proof of Theorem \ref{thm:consistPairwiseObj}, we have that our objective function in (\ref{eq:pwdiff}) is just two times the sample variance, $\text{var}_n(R).$ 
Note that therefore
\[\theta_{E,0} = \arg\min_{\theta_E} 2\text{var}_0(R(\theta_E,A_1,S_1))\] and  \[\theta_{E,n}=\arg\min_{\theta_E} 2 \text{var}_nR.\]
    Note that $var_0(\theta_E)$ is continuous in $\theta_E,$ differentiable, and has a unique minimum at $\theta_{E,0}.$
    Hence, by theory on the consistency of extremum estimators \citep{amemiya1985advanced}, if we can show that $\text{var}_n(R)$ converges \textit{uniformly} in probability to $\text{var}_0(R)$, we have that \[\theta_{E,n}\gop \theta_{E,0}.\]  

     We must show that $\text{var}_n(R)$ is Lipschitz. This, in conjunction with the facts that (1) $\theta_{E,0}$ is in a compact set $[0,1]$ and (2) we have pointwise convergence in probability, $\text{var}_n(R)\gop \text{var}_0(R)$, would allow us to conclude, by Theorem \ref{thm:consistPairwiseObj}, that we have, by \citet{newey1991uniform}, uniform convergence in probability of $\text{var}_n(R)$ to $\text{var}_0(R)$.

     Hence, we need only to show that $\text{var}_n(R)$ is Lipschitz.  By the Mean Value Theorem, we can show that  a function is Lipschitz by showing that its first derivative is bounded. We therefore examine the derivative of $\text{var}_n(R)$  below.
Note that %
{
\begin{align*}
&\text{var}_n(R) =\frac{1}{n}\sum_i (-((a_i-A^*)^2(\theta_Es_i + (1-\theta_E) (1-s_i)))\\&-\frac{1}{n}\sum_{k=1}^n -((a_k-A^*)^2(\theta_Es_k + (1-\theta_E) (1-s_k))       
)^2.
\end{align*}
}
We need to show that the derivative of the preceding display is bounded, which occurs if the derivatives of the summands are bounded. 
Differentiating the summand with respect to $\theta_E,$ we obtain 
{
\begin{align*}
 & 2(-((a_i-A^*)^2(\theta_Es_i + (1-\theta_E) (1-s_i)))\\&-\frac{1}{n}\sum_{k=1}^n  -((a_k-A^*)^2(\theta_Es_k + (1-\theta_E) (1-s_k))       
)\\
&\times 
-((a_i-A^*)^2(2s_i-1))
\\
&-\frac{1}{n}\sum_{k=1}^n -((a_k-A^*)^2(2s_k -1)),      
\end{align*}
}
which is bounded as long as $a_k$ is bounded, which holds by assumption. Since the derivative of $\text{var}_n$ is bounded, we have that $\text{var}_n$ is Lipschitz. By the preceding discussion, we therefore have that $\theta_{E,n}\gop \theta_{E,0}.$

\end{proof}

\subsection{Other statistical summaries}
\label{app:otherStatistics}
One might also consider comparing the average divergence from optimality for each group $k$ as 
\[
\frac{1}{n_E}\sum_{i\in I_E} (A_{k,i}-A^*)^2 - \frac{1}{n_C}\sum_{i\in I_C} (A_{k,i}-A^*)^2,
\]
 where $I_E$ are the indices of all exposed mice and vice versa for $I_C.$ This statistic depends on two statistics that use only half of the observations each, whereas $\theta_{E,n}$ is estimated using all observations.  Further, the magnitude of this statistic will depend on the scales of $A$ and $A^*,$ whereas $\theta_{E,n} \in [0,1]$ for any problem.    One might consider normalizing $d_n$ somehow, but these types of manipulations become ad-hoc. Also, the lack of an optimization objective makes it difficult to see what the statistic means in terms of the exposure, whereas with $\theta_{E,n}$, it is clear: $\theta_{0,E_n}$ is the exposed animal's tolerance for divergence from optimality, as it relates to the control, and it is the only statistic that solves the optimization in (\ref{eq:pwdiff}).
 Another statistic that does not require specification of the behavioral policy would just be a Wilcoxon test for the actions in the two groups, and although this does not assume a model, it is nonparametric, whereas our approach is still parametric, facilitating consistency arguments such as the ones made in Theorem \ref{thm:consistencyThetaEn} even under  mild assumptions on the nature of the action $A$.



\bibliographystyle{plainnat}
\bibliography{references}

\end{document}